# A new algorithm for fitting orbits of multiple-planet systems to combined RV and astrometric data

Joseph Catanzarite

Jet Propulsion Laboratory, California Institute of Technology

Joseph.H.Catanzarite@jpl.nasa.gov

**Abstract**

To fully determine a planet's orbit we must specify its orientation in three dimensional space. Astrometric measurements of the host star's reflex motion cannot distinguish the ascending node of the orbit from the descending node, so the longitudes of the periastron and of the ascending node are only determined modulo 180 degrees. To break this degeneracy, we need information about the star's reflex motion in the direction orthogonal to the sky plane, which is provided by radial velocity (RV) measurements.

If an orbit is fitted from combined RV and astrometric data, the orbit should be physically consistent with both data sets. The Keplerian orbit of a planet is a highly nonlinear function of seven parameters. The astrometric orbit problem can be partially linearized via transformation to four linear parameters (related four Thiele-Innes constants) plus three nonlinear parameters: eccentricity, period and periastron time. The RV orbit problem can be partially linearized via transformation to two additional linear parameters plus the same three nonlinear parameters. Unfortunately, the two linear parameters from RV are not linearly related to the four linear parameters from astrometry. Because of this difficulty, currently available algorithms for fitting combined RV and astrometric data to multiple-planet systems employ at least five nonlinear parameters per planet. We have developed a new algorithm for fitting orbits of multiple planet systems from combined data sets of radial velocity and astrometric measurements. The new algorithm satisfies the RV-astrometry consistency requirement, while using three nonlinear parameters per planet. We expect the reduction in nonlinearity to give the algorithm a significant advantage in computation speed over existing algorithms. In this work, we describe the algorithm, which has been validated in the context of a recent double-blind planet detection simulation study.

**Periodograms and detection**

The first step in the detection process is to inspect the periodograms of the RV data, the two-dimensional astrometric data, and the combined RV and astrometric data for 'significant' peaks, i.e. those that exceed a certain detection threshold. A two-axis periodogram which combines the astrometric data is defined as the sum of the power in the ecliptic longitude (E) and ecliptic latitude (N) directions (Catanzarite, 2006) (Unwin, 2008).

We set our detection threshold at the 99% confidence level, so that random noise could produce a comparable signal only 1% of the time. We say that the FAP (false alarm probability) for this threshold is 1%. For a given data set, the 1% FAP threshold will depend on the trial frequencies used in the periodogram; the more frequencies probed, the higher the threshold corresponding to a given FAP. For each data set, the 1% FAP threshold is derived from an ensemble of periodograms from a Monte Carlo simulation of Gaussian noise sampled at the same cadence; the periodogram should be sampled with the desired set of trial frequencies.

A three-axis periodogram combining the astrometric and RV data can also be defined as the sum of the normalized power in the N and E astrometric periodograms and the RV periodogram; in each case the power is normalized by the associated noise variance. The three-axis periodogram is useful for intermediate-period planets near the detection threshold, for which RV and astrometric data both provide information. In some cases, the signal from a planet which is not detected above the threshold in either the RV or two-axis astrometric periodogram, will register above the threshold on the three-axis periodogram.

We first looked for evidence of 'hot planets', using a set of trial frequencies that probed for periods down to a few days. For evenly sampled data, the Nyquist frequency is defined as half the sampling rate; signals of higher frequency will appear aliased at frequencies below the Nyquist frequency, and thus are undetectable. The sampling for the synthetic data in this exercise had realistic cadence, including an annual solar exclusion gaps and elsewhere randomly deviates by up to 10% from even sampling. This kind of time sampling effectively probes time scales much shorter than the average interval between samples, so that the concept of Nyquist frequency is not well-defined. If we define a pseudo-Nyquist frequency as half the frequency corresponding to the *average* time interval between samples, we do not expect it to be an upper limit to the frequencies that can be probed by the data. The astrometric data (100 samples over 5 years) has a pseudo-Nyquist frequency of 0.1 $yr^{-1}$, corresponding to a period of 36.5 days, and the RV data (150 samples over 15 years) has a pseudo-Nyquist frequency of 0.2 $yr^{-1}$, corresponding to a period of 73 days. Indeed in the double-blind test, our team was successful in correctly detecting 9 of the 10 hot planets with period < 0.1 years, with no false detections. The shortest detected period was under two days, corresponding to a factor of almost 40 above the pseudo-Nyquist frequency!

If no evidence of 'hot planets' was found, we backed down to a smaller set of evenly-spaced trial frequencies that extended only up to the pseudo-Nyquist limit; the 1% FAP level for this smaller set of frequencies was determined by Monte Carlo simulation, as described previously. Since there were fewer frequencies, the 1% FAP threshold for this smaller set of trial frequencies is lower. The most significant peak usually gives a very good initial guess for the period of the most significant planet in the data, which jump-starts the nonlinear fitting process. The exception is that for a long period planet (i.e. period near or longer than the duration of observations) the periodogram will have maximum power at the lowest frequency, corresponding to the duration of observations. In this case we set the initial guess of the period to about twice the duration of observations.

**Hierarchical orbit-fitting scheme**

The hierarchical scheme, in one form or another, was used by all four teams that participated in the double-blind multiple-planet fitting exercise (Traub, 2010). In our implementation, the first step is to use the periodogram to obtain an initial guess of the period of the most significant planet. The reflex motion orbit is then fitted and removed from the RV and astrometric data, after which the periodograms of the residuals are again inspected for significant peaks. If there is one, the most significant peak provides a starting guess for the period of the second planet, and the previously fitted parameters give starting guesses for the first planet, as well as the position and velocity offsets, the proper motion, and the parallax. The data is then fitted against a two-planet reflex motion model. This process is repeated iteratively, each time adding another planet to the model and updating all the previously fitted parameters, until there are no more significant periodogram peaks in the residual of the model and the data. The final iteration is a simultaneous full N-planet fit, including all the parameters. A more detailed breakdown of the approach is as follows:

a. Fit a model of parallax, proper motion, parallax and offsets to the astrometric data.

b. Remove this fitted model from the astrometric data, and remove a constant offset from the RV data.

c. Run the residuals through the astrometric and RV periodograms, respectively.

d. If a significant periodogram detection is observed in either or both periodograms, fit a one-planet model (N=1) including the 5 astrometric parameters and the RV offset parameter to the data, using the period from the periodogram as the initial guess.

e. Remove the fitted model from both the astrometric and RV data, and run periodograms on the residual

f. If second significant periodogram detection is observed in either or both periodograms, fit a N+1-planet model including the 5 astrometric parameters and the RV offset parameter to the data; in this fit, the parameters of the N previously fitted parameters are updated.

g. Remove the fitted model from the data, and run the periodogram on the residual

h. Continue, iteratively repeating steps (f) and (g), until no significant signal appears in the periodogram.

**Geometry of Keplerian orbits**

In this section we describe the geometric solution of a Keplerian astrometric reflex motion orbit (due to a single planet), from two-dimensional measurements of the star's trajectory on the sky, or from measurements of the star's radial velocity (van de Kamp, 1964). For a depiction of the geometry of a Keplerian orbit, see Figure 1 below. In the following description we assume that a solution of Kepler's equation i.e. period, periastron time, and eccentricity, is in hand.

We begin by stating some necessary definitions and conventions.

The **reference direction** is along the line of sight from the observer to the barycenter of the star-planet system. The star's radial velocity (RV) is the projection of the stellar velocity vector along this direction.

The **sky plane** is the plane orthogonal to the reference direction.

In the sky plane, the **+X** direction is along North and the **+Y** direction is along East.

The observed motion of the star due to its planetary companions is represented as time series **X (t)** and **Y (t)** in angular units on the sky (astrometric measurements) and **V (t)** in meters/sec (radial velocity measurements). In this discussion we will consider only the reflex motion of the star, neglecting position and RV offsets, proper motion, and parallax.

The **pericenter** is the point on the star's orbit at which it is closest to the barycenter (the center of mass of the star and the planet). The periastron time is the time at which the star is at the pericenter. It is chosen by convention to be the time at the first periastron passage during the observation window.

In the following analysis, we choose the **+Z** direction opposite to the reference direction. This is so that the **X**, **Y**, and **Z** directions form a right-handed Cartesian coordinate system, which we will call the **observer frame**. Note that since the RV observations are the projections of the stellar velocities onto the reference direction, we multiply them by -1 to bring them into the observer frame.

The **orbit plane** is the plane containing the reflex motion orbit. Without loss of generality, one can imagine viewing the orbit plane along a direction that is anti-parallel to the orbital angular momentum vector **L**, so that the star moves in a CCW orbit. Choose the **+x** direction in the orbit plane as the direction along the major axis of the orbit ellipse that contains the pericenter. Choose the **+z** direction along **L**, and choose the **+y** direction along the minor axis in the orbit plane so that the **x**, **y**, and **z** axes form a right-handed Cartesian coordinate system which we will call the **orbit plane frame**. The star's elliptical motion in the orbit plane, scaled to unit semimajor axis, is given by normalized dimensionless coordinates **x (t)**, **y (t)** which come from the solution of Kepler's equation (van de Kamp, 1964).

The **line of nodes** is the intersection between the orbit plane and the sky plane.

The **ascending node** of the star's reflex orbit is the point on the line of nodes at which the star crosses the plane of the sky moving away from the observer, so that its RV is positive in conventional coordinates.

The **longitude of the ascending node** $\Omega$ is the CCW angle in the sky plane from the North direction to the ascending node.

The **longitude of periastron** $\omega$ is the angle, in the orbit plane, along the direction of the star's orbit, from the ascending node to the pericenter.

The **inclination, I** of the orbit is the **CW** rotation angle about the ascending node direction (along the line of nodes) that brings the **+Z** axis into coincidence with the orbital angular momentum vector **L** of the reflex motion orbit. This definition is consistent with the convention that the observed orbit (i.e. the projection on the sky plane) is **direct** (CCW on the sky plane) for 0 < **I** < 90 degrees, and is **retrograde** (CW on the sky plane) for 90 < **I** < 180 degrees (Peale, 2003). We next identify the sequence of Euler rotations which transforms the observer frame to the orbit plane frame.

(1) Rotate by $\Omega$ about the **Z** axis. This brings the **X** axis into coincidence with the line of nodes, (with the ascending node along **+X**), which becomes the new **X'** axis.
(2) Rotate by **-I** (i.e. CW by I) about the new **X'** axis. This brings the **Z**-axis into coincidence with the orbital angular momentum vector **L,** which becomes the new **Z''** axis.

The new coordinate system looks down on the orbit plane along a direction anti-parallel to the angular momentum vector L, so that the motion in the orbit plane is CCW. All that remains to complete the transformation is

(3) Apply an azimuthal rotation of $\omega$ about **Z''**, bringing the **X'** axis into alignment with the **+x**-axis (the direction of the pericenter) in the orbit plane.

The sequence of three coordinate axis rotations described above transforms the coordinates from the observer reference frame to the orbit plane reference frame:

$$\begin{bmatrix} x(t) \\ y(t) \\ 0 \end{bmatrix} = \frac{1}{\alpha} R_Z(\omega) R_X(-I) R_Z(\Omega) \begin{bmatrix} X(t) \\ Y(t) \\ Z(t) \end{bmatrix}$$

In the above equation, $\alpha$ is the angular size of the semimajor axis on the sky in micro-arcseconds, and is the scale factor between the orbit plane ellipse and the observed ellipse.

Applying the sequence of inverse Euler rotations in opposite order transforms the orbit plane coordinates to the observer frame coordinates:

$$\begin{bmatrix} X(t) \\ Y(t) \\ Z(t) \end{bmatrix} = \alpha R_Z(-\Omega) R_X(I) R_Z(-\omega) \begin{bmatrix} x(t) \\ y(t) \\ 0 \end{bmatrix}$$

The Euler rotations we need are rotations about the X and Z axes. These are represented by the rotation matrices

$$R_Z(\theta) = \begin{bmatrix} \cos\theta & \sin\theta & 0 \\ -\sin\theta & \cos\theta & 0 \\ 0 & 0 & 1 \end{bmatrix} \text{ and } R_X(\theta) = \begin{bmatrix} 1 & 0 & 0 \\ 0 & \cos\theta & \sin\theta \\ 0 & -\sin\theta & \cos\theta \end{bmatrix}$$

We find

$$R_X(I)R_Z(-\omega) = \begin{bmatrix} 1 & 0 & 0 \\ 0 & \cos I & \sin I \\ 0 & -\sin I & \cos I \end{bmatrix} \begin{bmatrix} \cos\omega & -\sin\omega & 0 \\ \sin\omega & \cos\omega & 0 \\ 0 & 0 & 1 \end{bmatrix} = \begin{bmatrix} \cos\omega & -\sin\omega & 0 \\ \sin\omega\cos I & \cos\omega\cos I & \sin I \\ -\sin\omega\sin I & -\cos\omega\sin I & \cos I \end{bmatrix}$$

And

$$R_Z(-\Omega)R_X(I)R_Z(-\omega) = \begin{bmatrix} \cos\Omega & -\sin\Omega & 0 \\ \sin\Omega & \cos\Omega & 0 \\ 0 & 0 & 1 \end{bmatrix} \begin{bmatrix} \cos\omega & -\sin\omega & 0 \\ \sin\omega\cos I & \cos\omega\cos I & \sin I \\ -\sin\omega\sin I & -\cos\omega\sin I & \cos I \end{bmatrix}$$

$$= \begin{bmatrix} \cos\Omega\cos\omega - \sin\Omega\sin\omega\cos I & -\cos\Omega\sin\omega - \sin\Omega\cos\omega\cos I & -\sin\Omega\sin I \\ \sin\Omega\cos\omega + \cos\Omega\sin\omega\cos I & -\sin\Omega\sin\omega + \cos\Omega\cos\omega\cos I & \cos\Omega\sin I \\ -\sin\omega\sin I & -\cos\omega\sin I & \cos I \end{bmatrix}$$

So that

$$\begin{bmatrix} X(t) \\ Y(t) \\ Z(t) \end{bmatrix} = \alpha \begin{bmatrix} \cos\Omega\cos\omega - \sin\Omega\sin\omega\cos I & -\cos\Omega\sin\omega - \sin\Omega\cos\omega\cos I & -\sin\Omega\sin I \\ \sin\Omega\cos\omega + \cos\Omega\sin\omega\cos I & -\sin\Omega\sin\omega + \cos\Omega\cos\omega\cos I & \cos\Omega\sin I \\ -\sin\omega\sin I & -\cos\omega\sin I & \cos I \end{bmatrix} \begin{bmatrix} x(t) \\ y(t) \\ 0 \end{bmatrix}$$

Evidently the equations for $X$, $Y$, and $V$ are nonlinear in the standard orbit parameters.

Define geometric constants (related to the classical Thiele-Innes parameters):

$$A \equiv \alpha(\cos\Omega\cos\omega - \sin\Omega\sin\omega\cos I)$$
$$B \equiv \alpha(\sin\Omega\cos\omega + \cos\Omega\sin\omega\cos I)$$
$$F \equiv \alpha(-\cos\Omega\sin\omega - \sin\Omega\cos\omega\cos I)$$
$$G \equiv \alpha(-\sin\Omega\sin\omega + \cos\Omega\cos\omega\cos I)$$
$$C \equiv -a_*\sin\omega\sin I$$
$$H \equiv -a_*\cos\omega\sin I$$

In the above equations, $a_*$ is the semimajor axis of the stellar reflex motion orbit, in AU. Note that $a_* = \frac{\alpha}{\Pi}$, where $\alpha$ is the astrometric signature in uas and $\Pi$ is the parallax in uas. We can relate the stellar reflex motion orbit observables $X, Y$ and $Z$ to the orbit plane coordinates as follows:

$$\begin{bmatrix} X(t) \\ Y(t) \\ Z(t) \end{bmatrix} = \begin{bmatrix} A & F & -\sin\Omega\sin I \\ B & G & \cos\Omega\sin I \\ C & H & \cos I \end{bmatrix} \begin{bmatrix} x(t) \\ y(t) \\ 0 \end{bmatrix}$$

In the last equation X and Y are in micro-arcseconds (uas), while Z is in astronomical units (AU).

The motions in the observer's and orbit-plane frames are related as follows:

$$X(t) = Ax(t) + Fy(t)$$
$$Y(t) = Bx(t) + Gy(t)$$
$$V(t) = C\dot{x}(t) + H\dot{y}(t)$$

Here, $V \equiv \dot{Z}$ is the radial velocity of the stellar reflex motion, in the direction toward the observer, in AU per year, assuming the time units of the data are years.

Note that X and Y refer to the North and East directions on the sky, so that dX and dY are associated with motion along the directions of increasing ecliptic longitude and latitude, respectively.

The last equations are linear in the normalized orbit plane coordinates and their derivatives. This suggests that the orbital motions are separable into linear and nonlinear parts. The starting point is Kepler's equation for the orbit plane motion.

**Partial Linearization of the orbit problem**

Kepler's equation is

$$\mu(t-T) = E - e\sin E,$$

Where $\mu$ is the mean orbital motion $2\pi/P$ in radians/year, $P$ is the orbital period, $t$ is the time, $T$ is the periastron time, e is the eccentricity, and $E$ is the eccentric anomaly.

Given trial values of eccentricity, period and periastron time, we solve Kepler's equation (by an iterative method due to Danby (Danby, 1988)) for eccentric anomaly $E(t)$, so that the normalized orbit plane trajectory is

$$x(t) = \cos E(t) - e$$
$$y(t) = \sin E(t)\sqrt{1-e^2}$$

We can evaluate the time derivatives of $x(t)$ and $y(t)$ by differentiating Kepler's equation, obtaining

$$\mu = \dot{E} - e\dot{E}\cos E$$

So that

$$\dot{x}(t) = -\dot{E}\sin E = -\frac{\mu \sin E}{1 - e\cos E}$$
$$\dot{y}(t) = \sqrt{1-e^2}\,\dot{E}\cos E = \sqrt{1-e^2}\,\frac{\mu \cos E}{1 - e\cos E}$$

We can now write the equations for $X(t)$, $Y(t)$, and $V(t)$ in convenient matrix form

$$[X(t)] = \begin{bmatrix} \cos E(t) - e & \sin E(t)\sqrt{1-e^2} \end{bmatrix} \begin{bmatrix} A \\ F \end{bmatrix}$$

$$[Y(t)] = \begin{bmatrix} \cos E(t) - e & \sin E(t)\sqrt{1-e^2} \end{bmatrix} \begin{bmatrix} B \\ G \end{bmatrix}$$

$$[V(t)] = \begin{bmatrix} -\dfrac{\mu \sin E(t)}{1 - e\cos E(t)} & \sqrt{1-e^2}\,\dfrac{\mu \cos E(t)}{1 - e\cos E(t)} \end{bmatrix} \begin{bmatrix} C \\ H \end{bmatrix},$$

where each term involving $t$ is a column vector.

Each of these equations is nonlinear in the three parameters $e$, $P$, and $T$, and linear in two other parameters.

Given trial values of the nonlinear parameters, this system of equations is easily inverted by linear least squares to yield the six geometric constants $A$, $B$, $F$, $G$, $C$, $H$.

It is straightforward to modify these equations to include parallax, proper motion, position and radial velocity offsets in the linear model. In the Keplerian approximation, the reflex motion of a star with multiple planets is the superposition of the motions due to the individual planets, so the model can be easily extended to include any number of planets.

**Solution for the mass and orbital parameters**

Finally, we indicate how the four scale and orientation orbit parameters $\alpha$, $\omega$, $\Omega$, and $I$, as well as the planet mass M are obtained from the linear geometric parameters A, B, F, and G of the astrometric solution.

Following Elements of Astromechanics (van de Kamp, 1964), equations (6.9), define K, M and J as follows:

$$K = \frac{A^2 + B^2 + F^2 + G^2}{2},$$
$$M = AG - BF$$
$$J = \sqrt{K^2 - M^2}$$

Astrometric signature (in uas) and inclination (in radians) are then estimated as

$$\tilde{\alpha} = \sqrt{K + J}$$
$$\tilde{I} = \cos^{-1}\left(\frac{M}{K+J}\right)$$

As for $\omega$ and $\Omega$, we can solve only for their sum and difference.

$$(\omega + \Omega) = \tan^{-1}\left(\frac{B - F}{A + G}\right)$$
$$(\omega - \Omega) = \tan^{-1}\left(\frac{B - F}{A - G}\right)$$

These have range $[-\pi, \pi]$

Estimates of $\omega$ and $\Omega$ are therefore

$$\tilde{\omega} = \frac{\omega - \Omega}{2} + \frac{\omega + \Omega}{2}$$

$$\tilde{\Omega} = \frac{\omega + \Omega}{2} + \frac{\omega - \Omega}{2}$$

These also have range [-$\pi$, $\pi$]; evidently, they can only be determined modulo $\pi$ radians. Note also that if either one of $\Omega$ or $\omega$ is adjusted by $\pi$ radians, the other one must follow suit. So the astrometric data is consistent with two possible solutions, $\omega$, $\Omega$ or $\omega + \pi$, $\Omega + \pi$.

On the other hand, the RV data leads to an unambiguous solution for $\omega$, which when combined with the astrometric solution, determines $\Omega$.

The planet's orbital semimajor axis (in AU) is estimated from Kepler's 3rd Law:

$\tilde{a} = M_*^{1/3} \tilde{P}^{2/3}$, where $M_*$ is the mass of the star in solar units and $\tilde{P}$ is the estimate of the orbital period in years.

The semimajor axis of the stellar reflex motion orbit (in AU) is estimated as

$$\tilde{a}_* = \frac{\tilde{\alpha}}{\tilde{\Pi}}, \text{ where } \tilde{\Pi} \text{ is the estimate of parallax in uas.}$$

The center of mass equation gives the planet's mass $M_p M_\oplus = \frac{a_*}{a} M_* M_\odot$, where $M_p$ is the planet mass in units of $M_\oplus$, and $M_\odot$ and $M_\oplus$ are the masses of the Sun and Earth, respectively. The planet mass is therefore estimated as

$$\tilde{M}_p = \frac{M_*^{2/3}}{\tilde{P}^{2/3}} \frac{\tilde{\alpha}}{\tilde{\Pi}} \frac{M_\odot}{M_\oplus}$$

**Description of the joint RV and astrometry multiple-planet orbit fitting algorithm**

A star's apparent motion is the superposition of its reflex motion and its proper motion and parallax. We assume the reflex motion is Keplerian, so that it is the superposition of the reflex motions due to each of the planets.

Five astrometric parameters describe the non-gravitational part of the star's apparent motion: X and Y offsets, X and Y components of proper motion, and parallax. Parallax can be linearized, so this part of the model is linear.

The model for the astrometric reflex motion orbit can be partially linearized: for each planet there are three nonlinear parameters (eccentricity, periastron time, and period), and four linear parameters A, F, B, and G (related to the four astrometric Thiele-Innes constants) that are nonlinear functions of inclination, periastron angle, ascending node angle, and astrometric signature. Thus for an N-planet system, the astrometric reflex motion is modeled with 5 + 7N parameters, 3N of which are nonlinear.

The RV reflex motion model can also be partially linearized, each planet uses the same three nonlinear parameters (eccentricity, periastron time, and period), plus two additional linear parameters C and H (related to the $5^{th}$ and $6^{th}$ Thiele Innes constants) that are nonlinear functions of inclination and periastron angle. These parameters are not linearly related to A, B, F, and G. Additionally there is an RV offset, gamma. For an N-planet system, the RV reflex motion is modeled with 1 + 5N parameters, 3N of which are nonlinear.

When both astrometric and RV data are available, solving separately for RV and astrometric motion gives 6 + 9N parameters. This is sub-optimal, since the reflex motion orbit is completely specified in three-dimensional space by seven Keplerian parameters per planet. In effect there are two redundant parameters per planet, which sometimes give inconsistency between RV and astrometric solutions.

Accordingly, we have developed an efficient partial linearization scheme for multiple-planet orbit-fitting from RV and astrometric data that uses three nonlinear parameters and four linear parameters per planet. The scheme, described below, depends on forcing the solution to be consistent with both the RV and astrometric data.

We implemented this orbit-fitting scheme in MATLAB, using the **lsqnonlin** routine in MATLAB's optimization toolbox. We set up the problem as a constrained nonlinear optimization, specifying ranges for period (usually 0 to 30 years), periastron time (same as that of period) and eccentricity 0 to 1 (the upper limit can be reduced for cases in which suspected high-eccentricity bias is encountered). **lsqnonlin** starts with initial guesses we provide for periods of each detected planet (from the periodogram), eccentricity (zero), and periastron time (half the period). **lsqnonlin** operates in the standard way, evaluating the 'objective function' for the initial guesses, then iteratively adjusting the estimates until the 'objective function' is driven to a minimum. The objective function is the sum of the squared residuals of the model and the data. We provide details on its computation below.

The objective function is computed in the following way:

From the trial values of eccentricity, period and periastron time (provided by the nonlinear fitting routine, in this case, MATLAB's **lsqnonlin**) for each planet, solve Kepler's equation. This gives the normalized orbit-plane motion for each planet. Then invert the linear equations for the astrometric reflex motion, obtaining the four linear orbit parameters for each planet and the five astrometric non-reflex motion parameters. Compute the residuals of the fitted model with the X and Y astrometric data.

Next, invert the linear equation for RV reflex motion, obtaining the two linear parameters per planet associated with the radial velocity trajectory.

Astrometric data alone is two-dimensional; it cannot distinguish between the ascending and descending nodes, and therefore the periastron is only known modulo pi. RV data provides the third dimension, the information needed to specify the periastron angle. Since the astrometric-only solution determines Ω - ω, and Ω + ω, knowledge of the periastron also pins down the ascending node.

Combining astrometric-only and RV-only solutions uses nine fitted parameters to specify each planet orbit, and is thus sub-optimal.

We use the following simple scheme to obtain a joint solution that is consistent with both the astrometric and RV data, using the minimum of seven parameters per planet, three of which are nonlinear.

For each planet

1. Construct a model of the RV reflex motion due to this planet only (at the RV sampling times) from the astrometric-only solution.

$$[V_1(t)] = \left[ -\frac{\mu \sin E(t)}{1 - e \cos E(t)} \quad \sqrt{1-e^2} \frac{\mu \cos E(t)}{1 - e \cos E(t)} \right] \begin{bmatrix} C_1 \\ H_1 \end{bmatrix},$$

where $C_1 \equiv -\frac{\alpha}{\Pi} \sin \omega \sin I$, and $H_1 \equiv -\frac{\alpha}{\Pi} \cos \omega \sin I$

2. Construct an alternate model of the RV reflex motion due to this planet only (at the RV sampling times) from the astrometric-only solution, by adding 180 degrees to the periastron angle from the astrometric-only solution. This is the other possibility which is hidden in the astrometric solution.

$$[V_2(t)] = \left[ -\frac{\mu \sin E(t)}{1 - e \cos E(t)} \quad \sqrt{1-e^2} \frac{\mu \cos E(t)}{1 - e \cos E(t)} \right] \begin{bmatrix} C_2 \\ H_2 \end{bmatrix},$$

where $C_2 \equiv -\frac{\alpha}{\Pi} \sin(\omega + \pi) \sin I$, and $H_2 \equiv -\frac{\alpha}{\Pi} \cos(\omega + \pi) \sin I$

3. Generate a third model of the RV reflex motion due to this planet only (at the RV sampling times), this time from the solution of the RV-only problem.

   Compare models 1 and 2 against model 3, and select either model 1 or model 2, whichever has the lower residual with model 3. This choice determines which of the degenerate astrometric solutions $[\omega, \Omega]$ or $[\omega + \pi, \Omega + \pi]$ is most consistent with the RV data for this planet.

Assuming superposition of Keplerian orbits, summing the chosen RV model s over all the planets gives the model of the RV reflex motion for comparison with the mean-removed RV observations. We next compute the residual of this model with the mean-removed RV reflex motion data. The objective function for the nonlinear fit is the sum of the squared residuals of the X and Y and V models with the X, Y and V data, each normalized by their respective error. The normalized residuals are:

$$\delta X \equiv \frac{X_{model} - X}{\sigma_{ast}}$$

$$\delta Y \equiv \frac{Y_{model} - Y}{\sigma_{ast}}$$

$$\delta V \equiv \frac{V_{model} - (V - \bar{V})}{\sigma_{RV}}$$

The objective function is thus

$$F(\delta X, \delta Y, \delta V) = (\delta X)^2 + (\delta Y)^2 + (\delta V)^2.$$

Minimizing $F$ results in a solution that is consistent with both the astrometric and RV observations.

**Acknowledgements**

We are grateful to Tom Loredo, Barbara McArthur, Valeri Makarov, Mike Shao, and Chengzhing Zhai for helpful discussions. This work was carried out at the Jet Propulsion Laboratory, California Institute of Technology, under contract with NASA.

**Conclusion**

We have developed a new algorithm for fitting orbits of multiple planet systems from combined data sets of RV and astrometric measurements. The new algorithm re-formulates the orbit-fitting problem in terms of three nonlinear parameters per planet, taking advantage of partial linearization. Consistency is enforced between RV and astrometry. We expect the reduction in nonlinearity to give the algorithm a significant advantage in computation speed over existing algorithms. The algorithm has been validated in the context of a recent double-blind planet detection simulation study.

**References**

Catanzarite, J.H. et al. (2006). Astrometric Detection of Terrestrial Planets in the Habitable Zones of Nearby Stars with SIM PlanetQuest. *Publications of the Astronomical Society of the Pacific,* , 118, 1319.

Danby, J.M.A. (1988). *Fundamentals of Celestial Mechanics, 2nd edition.* Willmann-Bell.

Wright, J.T. and Howard, A.W. (2009). Efficient Fitting of Multi-Planet Keplerian Models to Radial Velocity and Astrometry, *Astrophysical Journal Supplement Series* 182, 205.

Lee, M.H. and Peale, S. J. (2003). Secular evolution of Hierarchical Planetary Systems *ApJ 592, 1201* .

Traub, W.A. (2010). In A. N. K. Gozdziewski (Ed.), *Extrasolar Planets in Multi-Body Systems: Theory and Observations. Ser. 42*, p. 191. Torun: EAS Publications.

Unwin, S. et al. (2008). Taking the Measure of the Universe: Precision Astrometry with SIM PlanetQuest. *Publications of the Astronomical Society of the Pacific* , 120, 38.

van de Kamp, P. (1964). *Elements of Astromechanics.* W.H. Freeman.

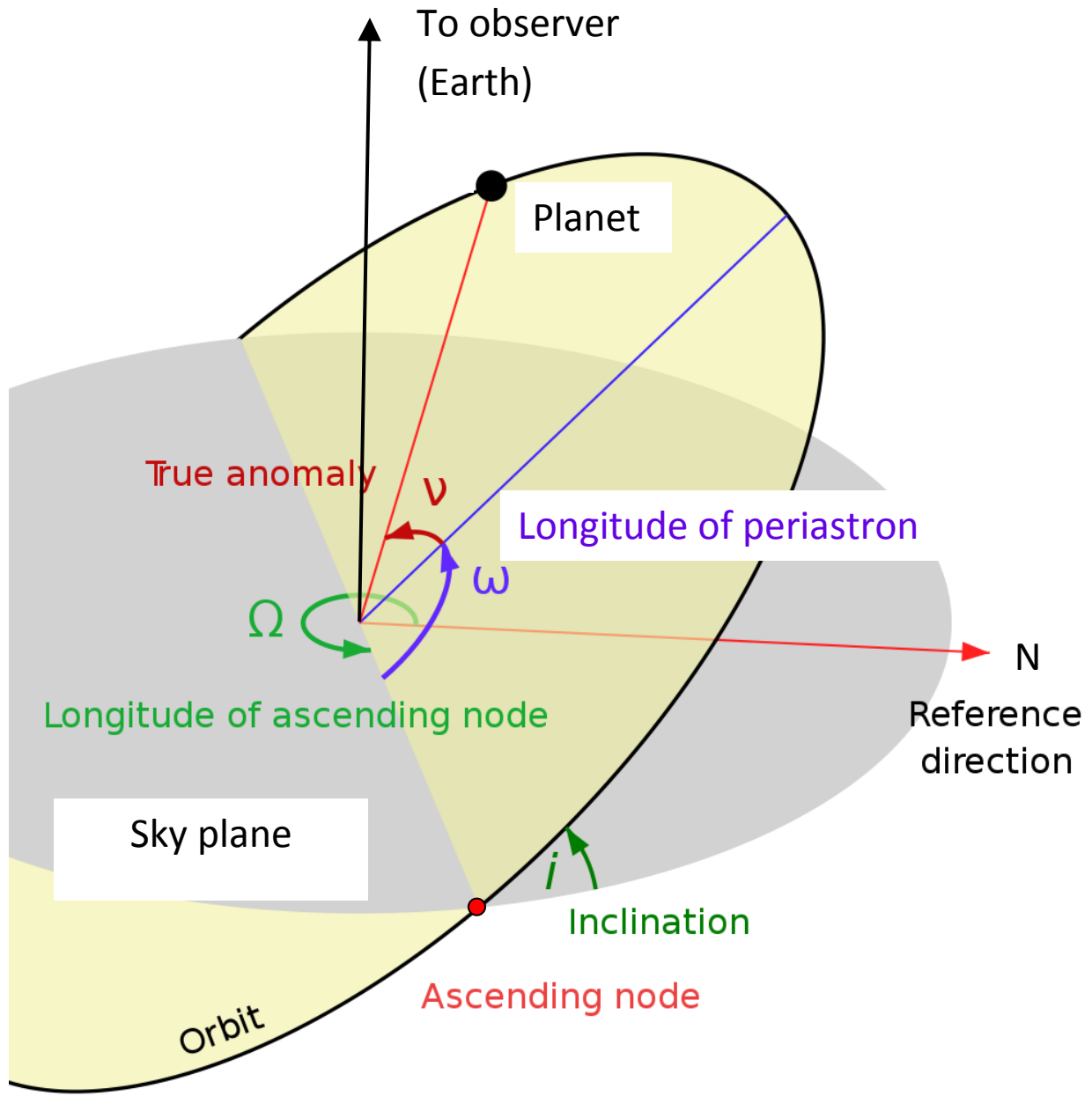

**Figure 1 Keplerian Orbit Geometry**

**(adapted from the image at http://en.wikipedia.org/wiki/Orbital_elements)**

15